\documentclass[journal]{IEEEtran}
\usepackage{cite}

\ifCLASSINFOpdf
  \usepackage[pdftex]{graphicx}
\else
  \usepackage[dvips]{graphicx}
\fi
\usepackage{hyperref}

\usepackage{booktabs}
\usepackage{multirow}
\usepackage{pifont}
\usepackage{microtype}
\usepackage{float}
\usepackage{enumitem}
\usepackage[table]{xcolor}

\begin{document}
%
\title{Acoustic Prompt Tuning: Empowering Large Language Models with Audition Capabilities}
%
%
%

\author{
    Jinhua~Liang,~\IEEEmembership{Student~Member,~IEEE,} 
    Xubo~Liu,~\IEEEmembership{Student~Member,~IEEE,}
    Wenwu~Wang,~\IEEEmembership{Senior~Member,~IEEE,}
    Mark~D.~Plumbley,~\IEEEmembership{Fellow,~IEEE,}
    Huy~Phan,~\IEEEmembership{Senior~Member,~IEEE,}
    Emmanouil~Benetos,~\IEEEmembership{Senior~Member,~IEEE}
\thanks{J.~Liang and E.~Benetos are with Centre for Digital Music (C4DM), Queen Mary University of London, UK. X.~Liu, W.~Wang, and M.~D.~Plumbley are with Centre for Vision, Speech and Signal Processing (CVSSP), University of Surrey, UK. H. Phan is with Meta Reality Labs, Paris, France.}
\thanks{The work does not relate to H.~Phan's position at Meta.}
\thanks{Email correspondence: jinhua.liang@qmul.ac.uk}}

\maketitle

\begin{abstract}
The auditory system plays a substantial role in shaping the overall human perceptual experience. While prevailing large language models (LLMs) and visual language models (VLMs) have shown their promise in solving a wide variety of language and vision understanding tasks, only a few of them can be generalised to the audio domain without compromising their domain-specific capability. In this work, we introduce \textbf{A}coustic \textbf{P}rompt \textbf{T}uning (APT), a new adapter extending LLMs and VLMs to the audio domain by injecting audio embeddings to the input of LLMs, namely soft prompting. Specifically, APT applies an instruction-aware audio aligner to generate soft prompts, conditioned on both input text and sounds, as the inputs to the language model. To mitigate data scarcity in the audio domain, a curriculum learning strategy is proposed by formulating diverse audio tasks in a sequential manner. Moreover, we improve the audio language model by using interleaved audio-text embeddings as the input sequence. In this improved model, zero constraints are imposed on the input format, thus it is capable of tackling diverse modelling tasks, such as few-shot audio classification and audio comparison. To further evaluate the advanced ability of the audio networks, we introduce natural language audio reasoning (NLAR), a new task that analyses two audio clips by comparison and summarisation. Experiments show that APT-enhanced LLMs (namely APT-LLMs) achieve competitive results compared to the expert models (i.e., the networks trained on the target datasets) across various tasks. We finally demonstrate APT's ability in extending frozen VLMs to the audio domain without fine-tuning, achieving promising results in audio-visual question and answering. Our code and model weights will be released at \url{https://github.com/JinhuaLiang/APT}.
\end{abstract}


%
\IEEEpeerreviewmaketitle

\section{Introduction}\label{sec:introduction}
%
%
%
%
\IEEEPARstart{A}{udio} plays a critical role in shaping human perception, enriching our understanding of the world beyond visual cues alone. Due to the prevalence of large language models (LLMs) and the advancement of visual language models (VLMs)~\cite{li_blip-2_2023,liu_visual_2023,dai_instructblip_2023,shukor_unified_2023,han_imagebind-llm_2023}, recent research primarily focus on visual and language modalities. Despite of the notable advances that have been made in image and video understanding, only a few of them~\cite{shukor_unified_2023,han_imagebind-llm_2023, Meng2023LLMTranscribeSpeech} have been extended to the audio domain.

To embrace more diverse modalities, more recent works have explored multi-modal tasks and heterogeneous data types. UniVAL~\cite{shukor_unified_2023} unified input/output format, model architecture, and training objective, and therefore, learned a shared encoder-decoder LLM with multi-modal curriculum learning. ImageBind-LLM~\cite{han_imagebind-llm_2023} adopted ImageBind, a cross-modal encoder bundling six modalities (including images) to a shared embedding space, and adapted the LLM with a frozen image encoder. While both approaches broaden the scope of visual LLMs, their reliance on specific architectures limits their adaptability for new modalities.

Following the VLM paradigm, several studies~\cite{Qwen-Audio, Hu2024WavLLMTR} have introduced audio-exclusive LLMs where a pair of an audio clip and a text token is used as an input for text generation. LTU~\cite{gong_listen_2023} bridged audio with language modalities by end-to-end fine-tuning on an instruction-based dataset.  
Qwen-Audio~\cite{Qwen-Audio} and SALMONN \cite{tang2024salmonn} expand such a framework into the speech and audio modalities via additional feature extraction using Whisper and BEATs encoders~\cite{10.5555/3618408.3619590,pmlr-v202-chen23ag}. 
Pengi~\cite{deshmukh_pengi_2023} learned from a large number of off-the-shelf datasets by using a multi-task framework. Still, Pengi is restricted to two domains (i.e., audio and language). These works cannot address tasks beyond the [audio, question, answer] format, e.g., few-shot audio classification~\cite{liang_adapting_2023}. To identify audio with a few labelled examples, Audio flamingo~\cite{kong2024audio} learned to predict the next text token from contextual information. However, these approaches are often restricted to fixed input formats, lacking flexibility for tasks beyond [audio, question, answer] structures. This constraint leads us to question: \textit{Can we adapt LLMs/VLMs to the audio domain by simply encoding sound clips as acoustic prompts?}

In this work, we introduce \textbf{A}coustic \textbf{P}rompt \textbf{T}uning (APT), an acoustic adapter that extends LLMs and VLMs to audio understanding and reasoning tasks by learning additional acoustic tokens (i.e., soft prompts) for the language model. Specifically, APT encodes audio clips into audio feature maps and employs an audio aligner (in Section~\ref{subsec:architecture}) to generate acoustic prompts. The acoustic prompts are conditioned on both the input instructions and the audio feature maps, aligning audio embeddings with the language modality. When training APTs, a curriculum learning strategy is adapted by formulating diverse audio tasks in a sequential manner. In addition to popular audio tasks, such as audio tagging (AT) and automated audio captioning (AAC), APT makes full use of publicly-available datasets where the model training is performed on three new tasks, namely query-based sound event detection, temporal event retrieval, and sound event counting, to learn fine-grained audio features. Moreover, we improve the audio language model framework by juxtaposing acoustic prompts with text embeddings. Rather than applying soft prompts as prefixes to the input texts, the improved framework exerts no constraints on the format of the input sequence. The APT-enhanced LLMs (APT-LLMs) interleave audio and text tokens to enhance in-context learning, preserve multimodal capabilities, and enable unified generative tasks. We have designed this approach to better align with our goals of making effective use of LLM capabilities while maintaining a generalizable and extensible framework. To further evaluate the advanced ability of our model, inspired by natural language visual reasoning~\cite{suhr-etal-2019-corpus}, we propose a new task referred to as natural language audio reasoning (NLAR) which is devised to distinguish, compare, and summarise two audio clips. Experiments on existing audio understanding tasks, including audio tagging, automated audio captioning, and few-shot audio classification, show that APT-LLMs achieve performance on par with those obtained by audio language models or domain-expert models. APT also yields promising performance on the proposed NLAR task, indicating its ability to compare and summarise over two different audio clips. Finally, quantitative studies are conducted to demonstrate the improved performance of APT over a VLM in an audio-visual question and answering (AVQA) task. Our contributions are summarised as below:
\begin{itemize}[parsep=-5pt,itemsep=5pt,labelsep=10pt]
    \item An acoustic adapter is introduced to extend LLMs and VLMs to the audio modality by soft prompting. To mitigate data scarcity issue, we introduce curriculum learning through a diverse set of tasks and a designed multi-stage training recipe. Leveraging the annotations in off-the-shelf databases, APT-LLM learns fine-grained acoustic embeddings from multiple audio tasks.
    
    \item APT formulates diverse audio tasks as a sequential task where generated text is conditioned on interleaved audio-text tokens. Without any constraints on the input format, APT-LLMs are not only able to solve different tasks according to the diverse instructions, but also to exploit the correlation among different audio clips in the same sequence. To the best of our knowledge, APT-LLM is the first audio-language model analysing beyond a single audio clip.
    
    \item A new audio analysis task, natural language audio reasoning, is proposed to distinguish, compare, and summarise two audio clips. Compared to existing audio tasks, this new task not only evaluates the ability of the model to understand an audio clip, but also requires the models to analyse the content of two recordings by comparison and summarisation. APT-LLM is then benchmarked on this task.
    
    \item A pretrained vision-language model, BLIP-2~\cite{li_blip-2_2023}, coupled with APT (namely APT-BLIP-2) is studied qualitatively and quantitatively on the audio-visual question and answering task~\cite{yang_avqa_2022}. APT-BLIP-2 can directly work with the visual modality without further fine-tuning, showcasing an efficient approach for extending multi-modal LLMs to a new modality.

\end{itemize}

\section{Related works}\label{sec:related_works}
\subsection{Multimodal language models}\label{subsec:multimodal_language_models}
In terms of recent advances, LLMs~\cite{touvron_llama_2023,chiang_vicuna_2023,openai_gpt-4_2023} have exhibited excellent comprehension and reasoning ability. Driven by the open-world knowledge in LLMs, a variety of visual language models have been proposed with different alignment methods to integrate image/video data with text tokens~\cite{alayrac_flamingo_2022,li_blip-2_2023,dai_instructblip_2023,zhang_llama-adapter_2023}. However, most of them are restricted to the visual domain. In other domains (such as audio), the development of such models is limited, largely due to the lack of training data and the presence of modality discrepancies. Recently, ImageBind-LLMs~\cite{han_imagebind-llm_2023} bridged the image encoder of ImageBind~\cite{girdhar_imagebind_2023}, a six-modality language model, with an LLM, and used visual tokens as soft prompts within the language model. UniVAL~\cite{shukor_unified_2023} unified the input/output, the architecture, and the training objective of multimodal LLMs and then devised a curriculum learning method for gradual exposure to the new modality. While both works adapted VLMs to other domains, they demand massive amount of multimodal data to train the overall networks from scratch. Instead, our proposed work investigates a domain-specific adapter that can extend any existing VLM/LLM to an additional modality (such as audio). 

\subsection{Audio language models}\label{subsec:audio_language_models} 
Following VLMs, some works built large audio pretrained models~\footnote{These are also referred to as ``audio foundation models'' in the literature} for sound-only tasks. SpeechGPT~\cite{zhang_speechgpt_2023} learned from a curated speech-text instruction dataset to analyse and generate spoken content in the audio domain. LTU~\cite{gong_listen_2023} was trained upon an open-ended dataset that contains 3.7M [audio, question, answer] tuples, with a perception-to-understanding curriculum. While LTU achieved a good audio analysis ability, it required a uniform input format as a triplet tuple. To work around this issue, Pengi~\cite{deshmukh_pengi_2023} proposed a multi-task framework where an audio language model is trained with off-the-shelf audio datasets prompted with different predefined questions. There are also audio language agents that apply LLM for audio manipulation \cite{liang2024wavcraft, liu2023wavjourney}.  Our work differs from these prior works in three ways: 1) Rather than being an audio-only language model, APT explores how to adapt existing VLMs and LLMs to the audio domain; 2) APT-LLM improves the curriculum learning by designing three new training tasks. By accessing existing datasets from different aspects, APT-LLM learns a fine-grained audio representation; and 3) APT-LLM re-frames the present input format, namely $<$audio, question, answer$>$, to allow audio and text to alternate in a sequence. In this way, APT-LLM is able to ingest more than one audio clip in a single feed-forward step, applying it to more audio tasks. To the best of our knowledge, APT-LLM is the first model that integrates in-context learning~\cite{von_oswald_transformers_2022} with curriculum learning.

\begin{figure*}[t]
    \centering
    \includegraphics[scale=1]{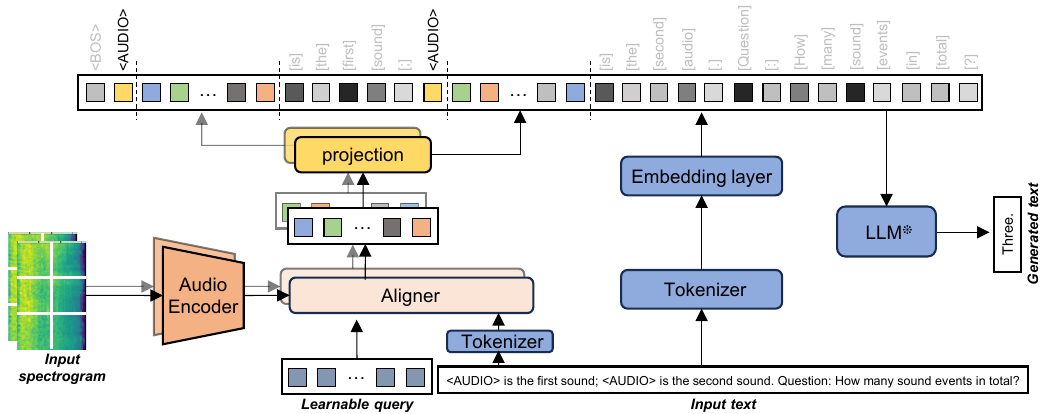}
    \caption{Illustration of the proposed APT-LLM. APT-LLM includes three components: an audio encoder, an audio aligner, and an LLM. The audio encoder extracts audio feature maps from input spectrograms. The audio aligner then projects each audio feature map to 32 acoustic embeddings according to the input text. These acoustic embeddings, together with the added embeddings of the audio token ``$[$AUDIO$]$'', are juxtaposed with text embeddings. The interleaved audio-text embeddings are fed into the LLM to generate the output text. APT-LLM can ingest multiple audio clips in a sequence and thus benefit from diverse tasks during training.}
    \label{fig.:lam}
    \vspace{-0.3cm}
\end{figure*}
\section{Acoustic Prompt Tuning} \label{sec:APT}
Current audio LLMs~\cite{gong_listen_2023,deshmukh_pengi_2023} learned to bridge audio with language by framing popular audio tasks (e.g., classification~\cite{10.1016/j.eswa.2023.121902,liang_adapting_2023} and captioning tasks~\cite{xu2022sjtu}) to the audio-conditioned text generation problem. Going beyond the [audio, question, answer] format, APT-LLM encodes multiple audio clips in one feed-forward process and juxtaposes them with text embeddings without any order constraint. This flexible training paradigm mitigates the need for high-quality annotations and massive databases, and thus reduces required computations. Moreover, juxtaposing audio clips with text enables APT-LLM to address more comprehensive audio analysis tasks, such as natural language audio reasoning. We first discuss the overall architecture of APT-LLM in Section~\ref{subsec:architecture}, and then elaborate the APT-LLM learning objective in Section~\ref{subsec:learning_objective} and the training recipe in Section~\ref{subsec:multi_task_learning}. In Section~\ref{sec:audio_reasoning_task}, we define the natural language audio reasoning task, a new task to evaluate the audio analysis ability of the proposed models.
 
\subsection{Architecture} \label{subsec:architecture}
The overall structure of APT-LLM is illustrated in Figure~\ref{fig.:lam}, with the main components being an audio encoder, an audio aligner, and a large language model. APT-LLM alternates audio clips with text tokens without any format constraints and thus benefits from task diversity and large-scale pretraining.

\textbf{Audio encoder: from spectrograms to feature maps.} We use Audio-MAE~\cite{huang_masked_2022}, a vanilla 12-layer transformer encoder that learns to reconstruct randomly-masked spectrogram patches during training, as the audio encoder. Rather than using the last layer fine-tuned for classification tasks, we apply the output feature map from the penultimate block of the Audio-MAE to encode the fine-grained patterns in audio. To obtain audio features, the input audio is processed in segments of 10 seconds. We zero-pad the clip when the input audio is shorter than 10 seconds. When the input is longer than 10 seconds, we clip the audio into 10-second segments and pad the last segment to 10 seconds if necessary. The Audio-MAE encoder takes the audio input in segments of 10 seconds and produces feature maps of shape (1024, 512). The feature maps are then sent into the audio aligner and then concatenated before fed into the large language model.

\begin{figure}[t]
    \centering
    \includegraphics[scale=1]{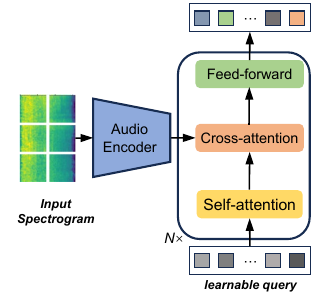}
    \caption{Structure of audio aligner in APT.}
    \label{fig.:aligner}
    \vspace{-0.3cm}
\end{figure}

\begin{table*}[t]
\centering
\caption{Curriculum learning adopted by APT-LLMs. APT-LLMs are trained with multiple tasks in different stages. ``\#Audio samples'' denote the number of audio clips in the dataset. Stage 0-2 denotes audio-text alignment, learning from single audio clips, and learning from multiple clips, separately.}
\label{tab.:multi-task_learning}
\resizebox{0.8\textwidth}{!}{%
\begin{tabular}{@{}lccccccc@{}}
\toprule
\multirow{2}{*}{Task} &
  \multicolumn{3}{c}{Training stages} &
  \multirow{2}{*}{Dataset} &
  \multirow{2}{*}{\#Audio samples} &
  \multirow{2}{*}{Durations} &
  \multirow{2}{*}{Setup} \\ \cmidrule(lr){2-4}
                                  & 0 & 1 & 2 &             &      &       &            \\ \midrule
Audio tagging                     & \ding{51}  & \ding{51}  & \ding{51}  & AudioSet    & 2M   & 5.8kh & train/test \\ \midrule
                                  &   &   &   & Wavcaps     & 400k & 7.6kh &            \\
Audio captioning                  & \ding{51}  & \ding{51}  & \ding{51}  & AudioCaps   & 39k  & 108h  & train/test \\
                                  &   &   &   & Clotho v2   & 7k   & 31h   &            \\ \midrule
Audio question and answering      &   & \ding{51}  & \ding{51}  & Clotho AQA  & 2k   & 12h   & train      \\ \midrule
Query-based sound event detection &   & \ding{51}  & \ding{51}  & AudioSet-SL & 81k  & 226h  & train      \\ \midrule
Temporal event retrieval          &   & \ding{51}  & \ding{51}  & AudioSet-SL & 81k  & 226h  & train      \\ \midrule
Sound event counting              &   & \ding{51}  & \ding{51}  & AudioSet-SL & 81k  & 226h  & train      \\ \midrule
Few-shot audio classification     &   &   & \ding{51}  & AudioSet    & 2M   & 5.8kh & train/test \\ \midrule
Natural language audio reasoning  &   &   & \ding{51}  & NLAR        & 0.2k & 1.2h  & train/test \\ \bottomrule
\end{tabular}%
}
\vspace{-0.3cm}
\end{table*}

\textbf{Audio aligner: from feature maps to audio tokens.} Audio aligner bridges the audio encoder to the frozen language model as shown in Figure~\ref{fig.:lam}. Figure~\ref{fig.:aligner} illustrates the structure of the audio aligner in APT. The audio aligner is constituted by $N$ transformer blocks, each of which contains self-attention, cross-attention, and feed-forward layers. 
The aligner processes a text prompt alongside acoustic features of shape (1024, 512) from the audio encoder to produce a fixed number of acoustic embeddings. Learnable query tokens and text embeddings from a BERT tokenizer are taken as inputs, with the query tokens extracting auditory information from audio feature maps via cross-attention.
For audio exceeding 10s, segments from audio embeddings are concatenated before being input into the large language model. Inspired by~\cite{li_blip-2_2023}, we adopted a Query-transformer (Q-former) architecture with four transformer layers. Unlike prior work~\cite{li_blip-2_2023}, our aligner specifically resamples variable-length audio embeddings into fixed acoustic embeddings, reducing computational complexity in downstream attention while filtering irrelevant information. The use of a BERT tokenizer aligns with the BERT-like transformer structure of our aligner. We set the number of learnable query tokens to 32 without additional specification.



\textbf{Large language model: from interleaved audio-text tokens to generated text.} As shown in Fig.~\ref{fig.:lam}, after encoding the whole text input with a dedicated text tokenizer, such as a LLAMA tokenizer~\cite{touvron_llama_2023}, the large language model extract audio and text embeddings separately. The language model generates text tokens sequentially by leveraging both the extracted input audio-text embeddings and its previously generated tokens to inform each prediction step. This allows the model to maintain position-aware contextual information throughout the generation process. We freeze all parameters in the language model during training. Different from existing works~\cite{li_blip-2_2023,gong_listen_2023}, we train a learnable audio token, ``$[$AUDIO$]$'', to identify audio modality from other modalities. A learnable token is prepended to each audio clip as a special token to indicate the position of audio tokens in the interleaved sequence. We find this token helps the language model to distinguish audio tokens from text tokens when interleaving them together.

\subsection{Learning objective}  \label{subsec:learning_objective}
In order to motivate our training framework, we first present the learning objective used in existing work~\cite{deshmukh_pengi_2023}. Let an audio-text pair in [audio, question, answer] format be referred to as $(a, t, g)$, where $a$, $t$, $g$ are the audio clip, input text, and output text, respectively, and $\mathbf{X}$ be the input sequential embeddings to the language model. To align the audio modality to the language modality, an audio encoder $\mathcal{A}$ and an audio aligner $\mathcal{M}$ project the audio $a$ into a sequence $\mathbf{X}_{\mathrm{audio}}$:
\begin{equation} \label{eqn.:extract_audio}
    \mathbf{X}_{\mathrm{audio}}\!=\!\mathcal{M}_{\theta}(\mathcal{A}_{\phi}(a), t),
\end{equation}
where $\phi$ and $\theta$ are the parameters of the audio encoder $\mathcal{A}$ and the aligner $\mathcal{M}$. The audio embeddings are used as a prefix to the input text embeddings as
\begin{equation} \label{eqn.:concat}
    \small{\mathbf{X}_{\mathrm{audio;text}}\!=\!\mathcal{C}(\mathbf{X}_{\mathrm{audio}}, \mathbf{X}_{\mathrm{text}}),}
\end{equation}
where $\mathcal{C}$ is a concatenating function. By combining Eqn.~\ref{eqn.:extract_audio} and Eqn.~\ref{eqn.:concat}, the audio embeddings can be obtained by
\begin{equation} \label{eqn.:concat_unfold}
    \small{\mathbf{X}_{\mathrm{audio;text}}\!=\!\mathcal{C}(\mathcal{M}_{\mathcal{\theta}}(\mathcal{A}_{\phi}(a), t), \mathcal{W}_{\psi}(t)),}
\end{equation}
where $\psi$ denotes the parameters of the word embedding layer $\mathcal{W}$ in the language model. Assuming the length of the concatenated embeddings $\mathbf{X}_{\mathrm{audio;text}}$ be $L$, the parameters of the audio LLM are optimised by measuring the probability distribution of the next token conditioned on its previous tokens:
\begin{equation} \label{eqn.:next_token_pred}
    \small{p(\mathbf{X}_{\mathrm{pred}}|\mathbf{X}_{\mathrm{audio}, \mathrm{text}})\!=\!\prod_{i=L+1}^{L+|g|}p_{\phi,\theta,\psi}(\mathbf{x}_i|\mathbf{X}_{\mathrm{audio;text},<i}; \mathbf{X}_{\mathrm{pred},<i}),}
\end{equation}
where $\mathbf{X}_{\mathrm{pred}}$ is the predicted sequence, and $\mathbf{x}_i$ is the $i$-th token of $\mathbf{X}_{\mathrm{pred}}$. In this way, prevailing LLMs are able to unify many audio-to-text tasks in a sequential manner. 

However, the [audio, question, answer] tuple format struggles to fully leverage rich context and structured information, hindering learning from contextual information. In contrast, the interleaved format improves in-context learning by exploiting structured information. For instance, a sequence like ‘(Input: [audio\_0], Output: [text\_0]), (Input: [audio\_1], Output: [text\_1]), (Input: [audio\_2], Output: ...)’ allows audio features to be directly associated with their corresponding text descriptions, facilitating more effective learning and comprehension.
We thereby propose a new learning framework in which interleaved audio-text embeddings are used as the LLM's input such that the model is able to leverage and learn from more diverse tasks during training. Let $\mathbf{a}$ and $\mathbf{t}$ be audio clips and input texts, and $g$ still be the output text. Assuming both $\mathbf{a}$ and $\mathbf{t}$ have $N$ different elements, the input audio-text pairs are denoted as $[(a^i, t^i)]_{i=1}^N$ where $a^i$ and $t^i$ are the $i$-th audio clip and input text, respectively. Eqn. (\ref{eqn.:concat}) can be re-written as
\begin{equation} \label{eqn.:interleave_abs}
    \mathbf{X}_{\mathrm{audio;text}}\!=\!\mathcal{I}(\mathbf{X}_{\mathrm{audio}}, \mathbf{X}_{\mathrm{text}}),
\end{equation}
where $\mathcal{I}$ is the function that alternates acoustic embeddings with text embeddings. More specifically, the interleaved embeddings can be obtained by
\begin{equation} \label{eqn.:interleave}
    \small{\mathbf{X}_{\mathrm{audio;text}}\!=\![(\mathcal{S}(a^{1}, t^{1})), T_{\psi}(t^{1}), \ldots, (\mathcal{S}(a^{N}, t^{N})), \mathcal{W}_{\psi}(t^{N})],}
\end{equation}
where $\mathcal{S(\cdot, \cdot)\!=\!\mathcal{M}_{\theta}(\mathcal{A}_{\phi}(\cdot), \cdot)}$. In this way, APT-LLM can integrate multiple audio clips in the input sequence, enabling itself to learn from more audio recognition tasks.

\subsection{Curriculum learning strategy}  \label{subsec:multi_task_learning}
With the uniform input/output format, APT-LLM is able to learn from a large variety of audio tasks and thus benefit from diverse training datasets. As shown in TABLE~\ref{tab.:multi-task_learning}, instead of passing through all training data directly, APT-LLM is trained through a three-stage curriculum:

\textbf{Audio-text alignment.} Before being coupled with an LLM, we pretrain the APT audio aligner to bridge the audio modality and the text modality. To this end, we freeze the audio encoder in the pipeline and optimise the parameters of the audio aligner with the audio-text pairs from AudioSet~\cite{gemmeke_audio_2017} and WavCaps~\cite{mei_wavcaps_2023}. During training, a fixed number of acoustic embeddings are learnt to extract relevant information from the audio feature maps according to the input text tokens. Following~\cite{li_blip-2_2023}, the audio aligner is learned with the triplet objectives: audio-text matching, audio-grounded text generation, and audio-text contrasting (see more details in Appendix~\ref{appendix:audio-text_alignment}).

\textbf{Learning from a single audio clip.} 
After APT has extracted acoustic embeddings according to the input text, the succeeding LLM learns to predict the next text tokens conditioned on both acoustic embeddings and the input text tokens. APT-LLM is thus trained with multiple tasks using various prompts (see more in Appendix~\ref{appendix:multitask_prompt}). In addition to existing audio tasks, namely audio tagging, automated audio captioning, and audio question and answering, we design three new tasks: (1) \textit{Query-based sound event detection} that aims to train a model to predict the onset and offset time of a specific sound event; (2) \textit{Temporal event retrieval} that is to recognise sound events occurred in a specific period, and (3) \textit{Sound event counting} that requires a model to count the frequency of a specific sound event in a recording. Instead of rendering datasets, we exploit the publicly-available AudioSet with strong labels~\cite{hershey_benefit_2021} using different prompts (see more in Appendix~\ref{appendix:multitask_prompt}). This curriculum learning strategy facilitates APT-LLM's learning from diverse datasets, including AudioSet~\cite{gemmeke_audio_2017}, WavCaps~\cite{mei_wavcaps_2023}, AudioSet with strong labels~\cite{hershey_benefit_2021}, Clotho~\cite{drossos_clotho_2020}, AudioCaps~\cite{kim_audiocaps_2019}, and Clotho-AQA~\cite{lipping_clotho-aqa_2022}.

\textbf{Learning from multiple audio clips.}
In addition to the aforementioned tasks, APT-LLM learns from two additional tasks by juxtaposing more than one audio clips with the input text. Specifically, few-shot audio classification and natural language audio reasoning are added to the curriculum learning framework in this stage. On the one hand, for the few-shot audio classification, APT-LLM predicts labels of sound events by exploiting the correlation between input audio clips. On the other hand, APT-LLM is required to compare and summarise two different sounds in the natural language audio reasoning task (see the following Section~\ref{sec:audio_reasoning_task}). Trained on these two tasks, APT-LLM learns to analyse beyond a single recording and answer questions as per input questions. We adopt AudioSet~\cite{gemmeke_audio_2017} and the proposed datasets for few-shot audio classification and natural language audio reasoning, respectively.

\subsection{Extending visual-language model to the audio domain} \label{subsec:audio-visual_modelling}
\begin{figure}[t]
    \centering
    \includegraphics[width=0.9\columnwidth]{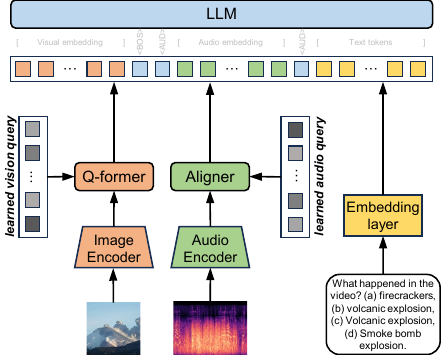}
    \caption{Illustration of APT-BLIP-2 for the audio-visual question answering task. Text prompts for the Q-former and Aligner are omitted for simplicity.}
    \label{fig.:avlm}
\end{figure}
APT was also investigated as an audio adapter for an existing VLM, BLIP-2~\cite{li_blip-2_2023}. BLIP-2 consists of a frozen image encoder, a 4-layer transformer, a projection layer, and a frozen Vicuna v1.1. Fig.~\ref{fig.:avlm} illustrates an example of APT-BLIP-2 for the audio-visual question answering task. 
Given raw audio $a$, an image input $v$, and a text query $t$, APT-BLIP-2 obtains audio embeddings $\mathbf{X}_{\mathrm{audio}}$ and visual embeddings $\mathbf{X}_{\mathrm{image}}$ with the audio aligner $\mathcal{M}$ and the Q-former $\mathcal{Q}$, respectively. In particular, the audio aligner downsamples audio features to create a sequence of audio embeddings by $\mathbf{X}_{\mathrm{audio}}=\mathcal{M}(\mathcal{A}(a), t)$. Meanwhile, the Q-former downsamples image features to a sequence of visual embeddings by $\mathbf{X}_{\mathrm{image}}=\mathcal{Q}(\mathcal{V}(v), t')$  where $\mathcal{V}$ denotes the image encoder and $t'$ is a fixed text prompt as in~\cite{li_blip-2_2023}. Before incorporating $\mathbf{X}_{\mathrm{image}}$, we concatenated audio and text embeddings in the format of “$<$BOS$>$ $<$AUDIO$>$ $\mathbf{X}_{\mathrm{audio}}$ $\mathbf{X}_{\mathrm{text}}$”. Finally, following BLIP2's approach~\cite{li_blip-2_2023}, we prepended visual embedding $\mathbf{X}_{\mathrm{image}}$ with the combined embeddings of audio and text, leading to the final format “$\mathbf{X}_{\mathrm{image}}$ $<$BOS$>$ $<$AUDIO$>$ $\mathbf{X}_{\mathrm{audio}}$ $\mathbf{X}_{\mathrm{text}}$”.

We refer to the APT-enhanced BLIP-2 as APT-BLIP-2. Of note, although we selected BLIP-2 as our backbone model, APT can be easily adapted to another vision language model.

\section{Natural language audio reasoning}\label{sec:audio_reasoning_task}
\subsection{Task formulation}\label{subsec:audio_reasoning_task}
One of the complex reasoning abilities of humans is to learn from different sounds, understanding what is happening in each audio and analysing the content of different audio by comparison and summarisation. However, existing audio tasks focus on analysing acoustic scenarios in individual recordings by recognising the sound events~\cite{kong_panns_2020} contained and/or retrieving their spatio-temporal information~\cite{politis_starss22_2022}. We thus propose natural language audio reasoning (NLAR), a new task where the model is required to answer questions by explicitly comparing or summarising two different audio recordings. TABLE~\ref{tab.:nlar_example} showcases an example of the NLAR task. A system takes two audio clips together with a free-form text query as input and is expected to answer the question by taking into consideration both audio recordings. Compared to existing audio tasks, the proposed audio comparison task features three notable differences:

\textbf{Comparison and summary between two audio clips}: This task requires a model to answer open-ended questions by comparing or summarising the content of two different audio clips. The model must first analyse the two audio clips as per the raised question separately and answer the question by taking into account the two audio inputs. An example of the audio comparison task can be found in TABLE~\ref{tab.:nlar_dataset}. 

\textbf{Diverse question types}: Questions for the natural language audio reasoning task assess diverse auditory aspects, such as the presence, the frequency, and acoustic features of sound events. Therefore, the model should not only identify the sound events in the recordings, but also retrieve relevant information as per the input question. 

\textbf{Effects of the chronological order}: Compared to existing audio tasks, e.g.,~\cite{li_blip-2_2023} and~\cite{gong_listen_2023}, the proposed audio comparison task emphasises the order of the audio recordings in a pair. In other words, the answer associated with the audio pair ``[Audio A, Audio B]'' could be different with the answer associated with  ``[Audio B, Audio A]'' when the questions are the same. In this way, we expect audio understanding models to be able to attend to different portions of the input sequence when the question is varied. 

By evaluating audio language models on the natural language audio reasoning task, we achieve more comprehensive assessment of the audio language models.
\begin{table}[t]
\centering
\caption{An example demonstrating APT-LLM's ability for audio reasoning. This task requires audio networks to comprehend recordings and reason across multiple recordings.}
\label{tab.:nlar_example}
\resizebox{\columnwidth}{!}{%
\begin{tabular}{@{}ll@{}}
\toprule
\multicolumn{2}{l}{Natural Language Audio Reasoning (NLAR) example: \textit{``Where is the sudden sound?''}} \\ \midrule
User         &  \\
             & \includegraphics[width=1.0\columnwidth]{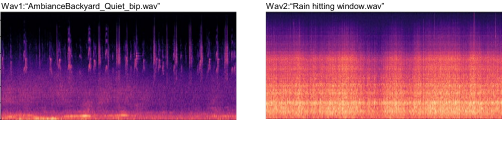} \\
             & Question: "Which recording has a more sudden and startling sound event?"             \\  \midrule
APT-LLM      & ``First."      \\
Ground truth & ``first"       \\ \bottomrule
\end{tabular}%
}
\vspace{-0.3cm}
\end{table}

\subsection{Data curation}\label{subsec:data_curation}
\begin{table}[t]
\centering
\caption{An example of rendering the NLAR dataset}
\label{tab.:nlar_dataset}
\begin{tabular}{@{}l@{}}
\toprule                                                                      
\begin{tabular}[c]{@{}l@{}}Based on the following two audio clips, generate 5 different questions that\\ must be derived by summarising or comparing both audios and is prohibited\\ to contain the information indicating its answer. The following information\\ is provided: the sound events appear in the audio clip, together with its\\ acoustic features, and corresponding onset and offset time stamps or fre-\\quency in the recordings. The answer should be either a binary one that\\ can be responded by `yes' or `no', or an open-ended one that can be reply\\ by a number or a single word.\\ \\ Audios:\\ \hspace{0.4cm} First: \\ \hspace{0.8cm} wood creaking (rustic, rhythmic, and creaky) {[}6 times{]};\\ \hspace{0.4cm} second: \\ \hspace{0.8cm} thunder (explosive, rumbling, and reverberating) {[}1 times{]}; \\ \hspace{0.8cm} rain (gentle, pitter-patter, rhythmic) {[}throughout the recording{]}.\\ Questions and answers:\\ \hspace{0.4cm} 1. are there 6 wood creaking and 2 thunder sounds in total? - no\\ \hspace{0.4cm} 2. how many wood creaking or thunder are there in total? - 7\\ \hspace{0.4cm} 3. does the second recording more likely to make one feel calm? - yes\\ \hspace{0.4cm} 4. in which recording the events are more frequent? - first\\ \hspace{0.4cm} 5. the frequency of wood creaking in the first recording is 6 times \\ \hspace{0.8cm} more than the frequency of thunder in the second one. - yes\\ \\ Audios:\\ \hspace{0.4cm} First: \\ \hspace{0.8cm} tapping glass (crisp, clear, and tingling sound) {[}9 times{]}; \\ \hspace{0.4cm} second: \\ \hspace{0.8cm} water (rapid, and draining water sound) {[}throughout the recording{]}.\\ Questions and answers:\\ \hspace{0.4cm} 1. are there 9 tapping glass sounds and 3 water sounds in total? - no\\ \hspace{0.4cm} 2. how many tapping glass or water sounds are there in total? - 10\\ \hspace{0.4cm} 3. does the second recording create a continuous sound throughout? - yes\\ \hspace{0.4cm} 4. in which recording are the sound events more repetitive? - first\\ \hspace{0.4cm} 5. does the second recording sound more dynamic compared to the \\ \hspace{0.8cm} first recording? - yes\\ \\ Audios:\\ \hspace{0.4cm} First: \\ \hspace{0.8cm} wood creaking (rustic, rhythmic, and creaky) {[}6 times{]};\\ \hspace{0.4cm} second: \\ \hspace{0.8cm} shower (droplets, soothingly cascading) {[}throughout the recording{]}.\\ Questions and answers: \end{tabular} \\ \bottomrule
\end{tabular}
\end{table}

Natural language audio reasoning (NLAR) is proposed to evaluate a model's ability in comprehending across audio clips as per the text instruction. We progressively render the NLAR dataset using the Clotho-AQA dataset~\cite{lipping_clotho-aqa_2022}, and every stage is described in the following:

\textbf{Data collection}: We manually filter out some audio samples from Clotho-AQA in terms of the quality of audio and the reliability of the annotations. To avoid data leakage, the test split of NLAR is collected from the test split of Clotho-AQA only.

\textbf{Data cleaning}: The annotation in the Clotho-AQA is noisy. For instance, annotators might not reach an agreement on the number of bird chirps in a recording. Therefore, we manually re-annotate the audio files by focusing on annotations where there is disagreement between annotators. We notice that it is infeasible to calculate the frequency of sound events in some cases, such as raining. In this case, we will annotate the presence of these activities throughout the recordings by using tags such as ``in the beginning'', ``in the middle'', ``in the end''  and ``throughout the recording''.

\textbf{Describing acoustic features}: We applied ChatGPT-turbo~\cite{openai_gpt-4_2023} to describe the acoustic feature of each sound event with the prompt ``describe the acoustic characteristic of \{SOUND\} precisely with a sentence less than 10 words'' where \{SOUND\} refers to the name of the sound event. We inspect each audio feature description to ensure they are comprehensible.

\textbf{Create audio-text pairs}: We applied ChatGPT-turbo to generate five question-answer pairs according to a) their associated sound events; b) the temporal information, and c) audio description of the two audio recordings.

\begin{table*}[t]
\centering
\caption{Zero-shot audio understanding performance (\%) comparison with audio language models. We group the methods in terms of their training strategy. ``\#Params.'' denotes the number of trainable parameters and ``\#Pairs'' represents the number of audio-text pairs. $\uparrow$ indicates the higher number, the better performance.}
\vspace{1em}
\label{tab.:existing_tasks_4}
\resizebox{2\columnwidth}{!}{%
\begin{tabular}{@{}llllcc@{}}
\toprule
Model & \#Params. & \#Pairs & AudioSet (mAP$\uparrow$) & AudioCaps (SPICE$\uparrow$) & Clotho (SPICE$\uparrow$) \\ \midrule
\textit{Audio-language models trained with the contrastive loss} &        &      &      &      &      \\
AudioCLIP~\cite{guzhov_audioclip_2022}                 & 30M    & 2M   & 25.9 & -    & -    \\
CLAP~\cite{elizalde_clap_2023}                         & 190M   & 128k & 5.8  & -    & -    \\ \midrule
\textit{One-for-all models for various audio tasks}              &        &      &      &      &      \\
Qwen-Audio~\cite{Qwen-Audio}
       & 640M    & - & 18.5 & - & 13.6 \\
LTU~\cite{gong_listen_2023}                            & 96M    & 5.7M & 18.5 & 17.0 & 11.9 \\
Pengi~\cite{deshmukh_pengi_2023}                       & $>$191M & 3.4M & -    & 18.2 & 12.6 \\
\rowcolor{lightgray!60}APT-LLM                          & 101M   & 2.6M & 14.7 & 17.1 & 11.6 \\ \bottomrule
\end{tabular}%
}
\end{table*}

\begin{table*}[t]
\centering
\caption{Performance (\%) comparison in automated audio captioning tasks. $\uparrow$ indicates the higher number, the better performance.}
\label{tab.:existing_tasks_5}
\begin{tabular}{@{}lcccccc@{}}
\toprule
\multirow{2}{*}{Model}                                          & \multicolumn{2}{c}{AudioCaps} & \multicolumn{2}{c}{Clotho} & \multicolumn{2}{c}{Weighted average}     \\ \cmidrule(l){2-7} 
 & SPICE$\uparrow$ & SPIDEr$\uparrow$ & SPICE$\uparrow$ & SPIDEr$\uparrow$ & \multicolumn{1}{c}{SPICE$\uparrow$} & \multicolumn{1}{c}{SPIDEr$\uparrow$} \\ \midrule
\multicolumn{7}{l}{\textit{Specialised systems trained with task-specific examples}}                                                                           \\
AT+CNN10~\cite{xu_investigating_2021} & 16.8         & -         & 11.5        & -       & 15.1          & -          \\
CNN-GPT2~\cite{kim_prefix_2023} & 16.7         & 43.8         & 11.1        & 21.5       & 14.9          & 36.7          \\
WSAC+PD~\cite{Kouzelis2023} & 17.3         & 40.3         & 12.3        & 24.7       & 15.7          & 35.3          \\
WavCaps~\cite{mei_wavcaps_2023} & 18.2         & 48.5         & 13.3        & 29.7       & 16.6          & \textbf{42.5} \\ \midrule
\multicolumn{7}{l}{\textit{One-for-all models for various audio tasks}}                                                                                        \\
APT-LLM                 & 19.1         & 40.2         & 13.2        & 24.8       & \textbf{17.2} & 35.3          \\ \bottomrule
\end{tabular}
\end{table*}

\textbf{Inspect the rendered data}: We manually checked the generated question-answer pairs using the annotations of the recordings and called ChatGPT again if the number of qualified pairs is lower than five. We dropped the generated pairs only if they contain some factual errors. We noticed that during data rendering, some answers to the questions could be unclear. We did not exclude them from the dataset as it is also important for models to learn what cannot be perceived from the audio. The curated NLAR dataset consists of 1722 audio-text pairs in the training split and 1640 pairs in the test split. 

\section{Experiments} \label{sec:experiments}
APT was first coupled with LLMs (i.e., APT-LLM) and evaluated as a general-purpose audio learner on a variety of existing audio-related benchmarks, including audio tagging, automated audio captioning, and few-shot audio classification. To further assess its ability in analysing two audio clips of interest, APT-LLM was further benchmarked on the natural language audio reasoning task. In addition to audio analysis, we tested and analysed (quantitatively and qualitatively) APT as a zero-shot adapter to BLIP-2~\cite{li_blip-2_2023,dai_instructblip_2023}, a state-of-the-art VLM.
\subsection{Experimental setup} \label{subsec:experiment_setup}
Our models were implemented based on the BLIP-2 framework~\cite{li_blip-2_2023}. We used Audio-MAE~\cite{huang_masked_2022} as the audio encoder in all APT models we developed. Considering that Audio-MAE only contains 100M parameters, we used a two-layer transformer as the aligner to bridge the audio and text domains. Without an explicit statement, we coupled APT with a pretrained LLM named Vicuna 7B v1.1~\cite{chiang_vicuna_2023} for evaluation. We tested APT-LLM with two close-ended datasets: AudioSet~\cite{gemmeke_audio_2017} and ESC-50~\cite{piczak_esc_2015}; and four open-ended datasets: Clotho~\cite{drossos_clotho_2020}, AudioCaps~\cite{kim_audiocaps_2019}, NLAR (in Section~\ref{sec:audio_reasoning_task}), and AVQA~\cite{yang_avqa_2022}. We did not evaluate APT models on new designed tasks, because there is no proper baseline to compare. Notably, we still trained APT-LLMs with those diverse objectives to build up a versatile audio learner.

Adam optimiser~\cite{kingma_adam_2015} was used for model training. We applied a warm-up strategy in the first 2000 steps and used a cosine linear learning rate in the succeeding steps. We trained the APT models using three NVIDIA A100 (40G) GPUs. The audio-text alignment pretraining and curriculum learning took 5 days each.

We used accuracy and mean average precision (mAP) as performance metrics to evaluate the audio systems on classification and QA datasets~\footnote{The current audio question and answering datasets include single words as the ground truth so we use classification metrics in the evaluation following previous works~\cite{lipping_clotho-aqa_2022,yang_avqa_2022}.} while using SPICE \cite{Anderson2016SPICESP} and SPIDEr \cite{Liu2016ImprovedIC} on automated audio captioning datasets.

\subsection{Comparison with existing approaches} \label{subsec:comparison_with_existing_approaches}
We compare APT-LLM against the state-of-the-art specialised systems, i.e., the networks trained with task-specific data~\cite{mei_wavcaps_2023,xu2022sjtu,kim_prefix_2023,Kouzelis2023,liang_learning_2022,liang_adapting_2023}, and audio-language models~\cite{guzhov_audioclip_2022,elizalde_clap_2023,gong_listen_2023,Qwen-Audio,deshmukh_pengi_2023} on existing tasks, including audio tagging, automated audio captioning, and few-shot audio classification.

\textbf{Audio tagging} requires models to predict classes of test samples from a predefined label set. We evaluated the models on the AudioSet dataset~\cite{gemmeke_audio_2017}. During inference, APT-LLM was prompted using the sentence ``\textit{Summarize the audio with key words.}'' Since APT generates free-form texts directly, we used the APT text encoder pretrained in the stage 1 to encode generated answers and the given class names to text embeddings. Afterwards, cosine similarity is calculated as the classification probability.
Consistent with the findings in previous work~\cite{gong_listen_2023}, TABLE~\ref{tab.:existing_tasks_4} shows a performance gap between audio language models and task-specific models. This is expected since the latter model address the classification task as a close-set problem (as opposite to an open-set problem), with much lower complexity than open-set problem where models need to search across the entire word embedding space. In addition, we found that the performance of the text encoder greatly impacts the classification result when evaluating the generated answers. This finding can be attributed to the idea that word embeddings for different classes should remain sufficiently sparse when calculating their distance from the embeddings of the generated answer.

\textbf{Audio captioning} is the task where models are supposed to generate free-form description according to an input recording. The sentence ``\textit{Describe the audio clip concisely}.'' is applied as the input prompt. To mitigate the vocabulary gap between the datasets and the language model, which leads to a lower score of pretrained audio-language models on the language-related metrics, we fine-tune APT-LLM for two epochs on the training split of AudioCaps~\cite{kim_audiocaps_2019} and Clotho~\cite{drossos_clotho_2020} datasets and compare it with the captioning models trained on both tasks. As shown in TABLE~\ref{tab.:existing_tasks_5}, APT-LLM achieves the best weighted average SPICE score of 0.172 while yielding a comparable SPIDEr score of 0.353 against task-specific systems on automated audio captioning datasets.

\begin{table}[t]
\centering
\caption{Accuracy (\%) of various methods on ESC-50 in the few-shot settings.}
\label{tab.:few_shot}
\begin{tabular}{llc}
\toprule
                                          & \multicolumn{2}{c}{Accuracy$\uparrow$} \\ \hline
                                          & 5-way          & 12-way         \\ \cline{2-3}
\multicolumn{3}{l}{\textit{Specialised systems trained with task-specific examples}} \\
ProtoNet~\cite{snell_prototypical_2017}   & 88.2          & 77.7          \\
MatchNet~\cite{vinyals_matching_2016}     & 86.8          & 71.8          \\
HPN~\cite{liang_leveraging_2022}          & 88.7          & 78.7          \\ \midrule
\multicolumn{3}{l}{\textit{Audio language models trained with contrastive learning}} \\
TIP-adapter~\cite{zhang_tip-adapter_2022} & 97.5          & 95.6          \\
Treff adapter~\cite{liang_adapting_2023}  & 98.5          & 96.3          \\ \midrule
\multicolumn{3}{l}{\textit{One-for-all models for various audio tasks}} \\
APT-LLM                                & 91.0          & 54.2          \\ \bottomrule
\end{tabular}
\end{table}

\textbf{Few-shot audio classification} aims to classify test audio clips using labelled audio examples. Models were evaluated in the $N$-way $K$-shot problem~\cite{liang_learning_2022,liang2024mind} where: (1) there are $N$ classes in the classification task, and (2) each class contains $K$ different audio examples. Following previous works~\cite{vinyals_matching_2016}, in this task, we tested APT-LLM in the 5/12-way 5-shot settings where there are 5/12 types of sound events to classify and each type contains 5 audio examples. In our free-form query design, we prompt the few-shot classification question by adding the query audio clip together with the input text ``\textit{This is a sound of}'' to the end of the sequence of labelled audio examples and their corresponding label texts. We implemented the same evaluation protocol to all few-shot learners for a fair comparison. As shown in TABLE~\ref{tab.:few_shot}, APT-LLM outperforms the task-specific models~\cite{snell_prototypical_2017,vinyals_matching_2016,liang_leveraging_2022} in the 5-way 5-shot setting while having a competitive performance compared to CLAP Adapters~\cite{zhang_tip-adapter_2022,liang_adapting_2023}. In the 12-way 5-shot problem, however, we can observe a performance degradation of APT.

\begin{table}[t]
    \centering
    \caption{Benchmarking APT for accuracy~(\%) on the natural language audio reasoning task.}
    \label{tab.:audio_reasoning}
    \begin{tabular}{lc}
    \toprule
    Model               & Accuracy$\uparrow$         \\ \hline
    AAC+ChatGPT & 27.9 \\
    APT-LLM & 62.9              \\
    APT-LLM$_{1.5}$  & \textbf{63.8}     \\ \bottomrule
    \end{tabular}
    \vspace{-0.2cm}
\end{table}

\begin{table}[h]
    \centering
    \caption{Accuracy~(\%) comparison between different modalities in audio-visual learning.}
    \label{tab.:av_learning}
    \begin{tabular}{@{}lcc@{}}
    \toprule
    Model                         & Modal       & Accuracy$\uparrow$         \\ \midrule
    BLIP-2~\cite{li_blip-2_2023} & Video-only  & 42.9             \\
    APT-LLM             & Audio-only  & 27.7                  \\
    APT-BLIP-2                  & Audio-video & \textbf{59.7}    \\ \bottomrule
    \end{tabular}
    \vspace{-0.2cm}
\end{table}

We have observed some similar experimental findings in previous LLM research~\cite{jin2024longcontextllmsmeetrag,li2024longcontextllmsstrugglelong}. One plausible hypothesis, as per the conclusion in~\cite{li2024longcontextllmsstrugglelong}, is that the distribution of examples within prompts impacts model performance. In contrast, the gradient-based optimization methods can avoid such issues in the few-shot setting. Additionally, it should be noted that while APT-LLM was trained with 4-way 1-shot tasks, it can generalise to other few-shot settings, suggesting that APT-LLM learns to act as a few-shot classifier rather than memorising the expected answers. More discussion on few-shot learning can be found in Appendix~\ref{appendix:few-shot}.

\subsection{Evaluation on natural language audio reasoning} \label{subsec:evaluation_on_nlar}

Since APT-LLM is able to ingest multiple audio clips in a single feed-forward process, we investigated APT-LLM with natural language audio reasoning for which a model is expected to distinguish, compare, and summarise two audio clips (see Section~\ref{sec:audio_reasoning_task}). To the best of our knowledge, there is no previous work evaluating model ability to analyse more than one recording. We thus contrast APT-LLM to a cascaded model which connects an automated audio captioning model~\cite{xu2022sjtu} with ChatGPT-4o~\cite{openai_gpt-4_2023}. TABLE~\ref{tab.:audio_reasoning} compares different methods on the NLAR dataset. APT-LLM$_{1.5}$ denotes the combination of our APT models and Vicuna v1.5. It can be observed that APT-LLM$_{1.5}$ achieves 63.78\% mAP score, outperforming the baseline by a large margin. This result suggests that APT-LLM is able to not only recognise the content in an audio clip but also analyse more than one audio recording by comparison and summarisation. It is worth noting that there is marginal improvement when upgrading Vicuna from v1.1 to v1.5, indicating the performance of language models is not the bottleneck in this task, at least for the two used in our study.

\subsection{Evaluation on zero-shot audio-visual tasks} \label{subsec:evaluation_on_audio-visual_tasks}
APT-BLIP-2, together with other multimodal language models, was investigated on an audio-visual question and answering dataset where models are expected to choose one out of four options by using both audio and video modalities. We tested APT-BLIP-2 on the subset of the AVQA dataset~\cite{yang_avqa_2022} as many video links associated with the AVQA test segmentation were no longer available on the internet at the time of the experiment. As shown in TABLE~\ref{tab.:av_learning}, APT-BLIP-2 yielded a better performance than video-only and audio-only counterparts, indicating the adaptation to the audio domain benefits models' learning from the content of video. 

\subsection{Ablation study} \label{subsec:ablation}
\begin{table}[t]
    \centering
    \caption{SPICE score (\%) comparison by ablating the architecture of audio aligner.}
    \label{tab:ablate_agligner}
    \begin{tabular}{ccc}
       \toprule
        & AudioCaps & Clotho \\
       \midrule
       Linear & 6.54 & 7.49 \\
       Transformer  & 17.1 & 11.6 \\
       \bottomrule
    \end{tabular}
\end{table}

\begin{table}[t]
    \centering
    \caption{SPICE score (\%) comparison by ablating the choice of audio encoder.}
    \label{tab.:ablate_encoder}
    \begin{tabular}{cccc}
    \toprule
                            & w/ aligner & AudioCaps & Clotho \\ \midrule
    CLAP                    &            & 12.93     & 7.78    \\
    MAE                     &            & 6.01      & 3.06    \\
    MAE                     & \ding{51}  & 15.83     & 11.30   \\ \bottomrule
    \end{tabular}%
\end{table}

\begin{table}[t]
\centering
\caption{SPICE score (\%) comparison by ablating training tasks involved.}
\label{tab.:ablate_accat}
\begin{tabular}{@{}ccccc@{}}
\toprule
AAC       & AT      & AudioCaps & Clotho & Weighted average \\ \midrule
\ding{51} &         & 13.69     & 11.37   & 12.95   \\
          & \ding{51} & 7.64      & 2.89    & 6.12    \\
\ding{51} & \ding{51} & 15.83     & 11.30   & 14.38   \\ \bottomrule
\end{tabular}%
\end{table}

\begin{table}[t]
    \centering
    \caption{Zero-shot SPICE score (\%) of APT-LLM on automated audio captioning datasets by ablating the curriculum learning during training.}
    \label{tab.:ablate_curriculum}
    \begin{tabular}{ccc}
       \toprule
        & AudioCaps & Clotho \\
       \midrule
       Single-stage & 14.5 & 10.7 \\
       Multi-stage  & 17.1 & 11.6 \\
       \bottomrule
    \end{tabular}
\end{table}

\begin{table}[t!]
    \centering
    \caption{Accuracy (\%) comparison by ablating NLAR tasks during training.}
    \label{tab.:ablate_nlar}
    \begin{tabular}{ccc}
       \toprule
        & w/o NLAR & w/ NLAR \\
       \midrule
       Accuracy & 16.4 & 63.8 \\
       \bottomrule
    \end{tabular}
\end{table}

TABLE~\ref{tab:ablate_agligner} compares the performance of the proposed audio aligner against a simple linear layer on AudioCaps and Clotho datasets. For the comparison method, following~\cite{gong_listen_2023} implementation, we used a linear layer to connect audio encoder and LLM and trained this method with the datasets in Table~\ref{tab.:multi-task_learning}. This result indicates that the transformer architecture significantly outperforms the linear architecture on both tasks, achieving a SPICE score of 17.1 on AudioCaps and 11.6 on Clotho, compared to the linear model’s scores of 6.54 and 7.49, respectively. This experiment indicates that the transformer structure is highly effective for capturing complex audio patterns, resulting in substantially improved performance.

TABLE~\ref{tab.:ablate_encoder} compares the performance of two different audio encoders, MAE and CLAP, on the AudioCaps and Clotho datasets. All systems are trained on automated audio captioning and audio tagging datasets for simplicity. Since the MAE model was trained for spectrogram reconstruction, the length of output features is much longer than that of our audio aligner and the CLAP model. We evaluate two choices of downsampling approaches, mean average pooling in~\cite{gong_listen_2023} and learnable downsampling, to match the input requirement of the subsequent language model. The results show that the CLAP encoder significantly outperforms the MAE encoder with mean average pooling on both tasks, with SPICE scores of 12.93 on AudioCaps and 7.78 on Clotho while underperforms that with the learnable downsampling module. One possible reason for this disparity is that the downsampling approach will lead to a loss of audio information, thus impacting MAE's performance. We applied MAE as the audio encoder for APT because MAE preserves fine-grained information from audio and the audio aligner is expected to learn to extract the useful information by downsampling the audio features. We will leave the choice of the audio encoder for future exploration.

TABLE~\ref{tab.:ablate_accat} and TABLE~\ref{tab.:ablate_curriculum} assess the impact of curriculum learning by ablating the diverse set of tasks and multi-stage training recipe separately. TABLE~\ref{tab.:ablate_accat} shows that multi-task learning, training on both AAC and AT, yields the best performance on the downstream task, achieving 14.38\% of weighted average SPICE score on AudioCaps and Clotho datasets, outperforming single-task training on either AAC or AT alone. Specifically, the AT-only model exhibits significant performance degradation, with scores of 5.27\% weighted average SPICE score on AudioCaps and Clotho, suggesting the limitations of single-task pretraining on a task that may not align well with other downstream objectives, leading to a less effective learning. In contrast, multi-task learning enhances the generalisation ability of APT-LLM by leveraging the heterogeneity across different tasks.

TABLE~\ref{tab.:ablate_curriculum} shows the zero-shot performance of APT-LLM on automated audio captioning datasets by ablating the curriculum learning during training. Multi-stage learning improves the performance of APT-LLM by 2.6\% of SPICE on AudioCaps and 0.9\% of SPICE on Clotho, which indicates the effectiveness of the proposed curriculum learning.

TABLE~\ref{tab.:ablate_nlar} compares model performance, measured by accuracy, when trained with and without NLAR tasks. The results show a substantial performance improvement when NLAR is included during training, with accuracy increased from 16.4 (without NLAR) to 63.8 (with NLAR). The result suggests that incorporating NLAR tasks is essential for achieving high accuracy, especially when the model is evaluated on tasks closely related to NLAR.

\begin{figure}
    \centering
    \includegraphics[width=\columnwidth]{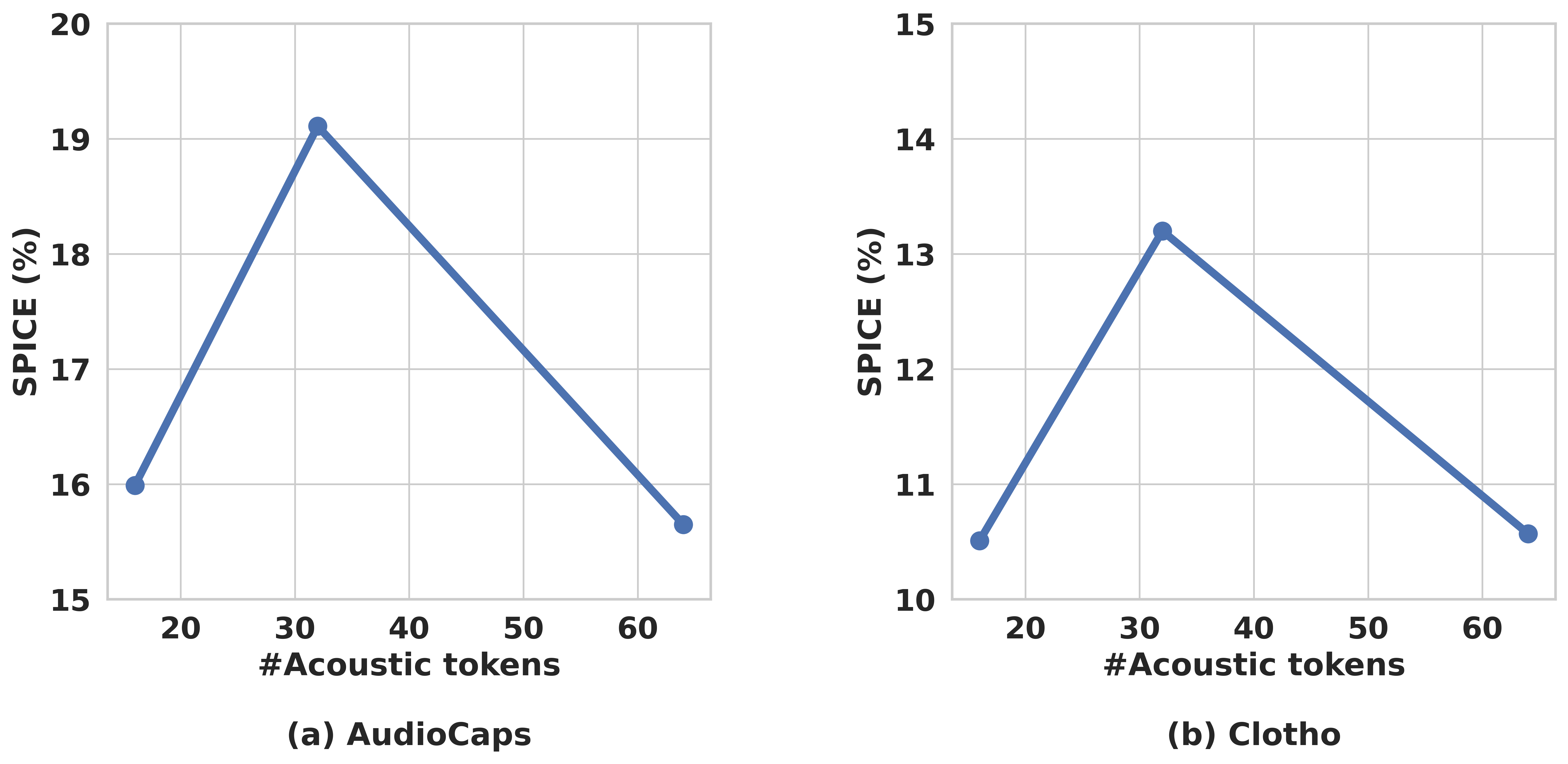}
    \vspace{-0.6cm}
    \caption{Performance comparison on automated audio captioning datasets, AudioCaps and Clotho, by varying the number of acoustic tokens.}
    \label{fig:ablate_token_num}
\end{figure}

We evaluated how the choice of token length impacts on the performance of the proposed method. The results are shown in Fig.~\ref{fig:ablate_token_num}. We can see that APT with 32 learnable acoustic tokens achieved the best performance compared to configurations with 16 and 64 acoustic tokens.

\section{Limitations} \label{sec:limitations}
The proposed method could be further improved in the following aspects. First, we devised APT to align acoustic embeddings with text embeddings obtained by language models. Word embeddings may change when a different language model is used, even if their architectures remain the same. As a result, each language model may need a dedicated APT aligner for adaptation. Second, APT-LLM was not trained with instruction-based datasets. Consequently, its ability is limited when responding to out-of-domain questions, which are not included in the training set. Finally, we focused the training and experimentation of APT-LLM on general-purpose audio understanding tasks; thus its capability has yet to be expanded to the realm of speech and music audio.

\section{Conclusions} \label{sec:conclusions}
We proposed APT, a general-purpose acoustic adapter that extends LLMs/VLMs to the audio domain. We showed that LLMs coupled with APT are versatile audio learners that not only achieved a competitive performance across various audio analysis tasks but also are capable of in-context learning~\cite{dai_why_2022} when fed with a few labelled examples. We also benchmarked APT-LLM's ability of audio comparison and summary using the natural language audio reasoning task, a new task that requires a model to distinguish, compare, and summarise two different audio clips. Last but not least, it is evident from our study on audio-visual learning that encoding sound clips as word tokens is helpful to adapt LLMs/VLMs to the audio domain. 

Future works can extend audio language models to jointly model the mixture of audio, music, and speech as they are complimentary for audio content analysis, especially in complicated real-world scenarios \cite{zhang2024llaqoassessment}. This could involve the choices of the audio encoder in an audio-language model. Also, as suggested in~\cite{gong_listen_2023}, instruction tuning is promising to align audio analysis with human perception by exploiting the capacity of large language models. It is also interesting to investigate how audio language models handle errors from in-context learning. Finally, scaling up audio reasoning and summary datasets is essential to benchmark audio language models on more complex tasks. One intriguing direction is to generate reliable annotations by using the existing fine-grained dataset, such as AudioSet strong-label, and large language models, e.g., ChatGPT. Another possible approach is to synthesise high-quality audio conditioned on target question-answer pairs by using generative models.

\section*{Acknowledgment} \label{sec:acknowledgment}
This work was supported by the Engineering and Physical Sciences Research Council [grant numbers EP/T518086/1 and EP/T019751/1]. E. Benetos was supported by a RAEng/Leverhulme Trust Research Fellowship [grant numbers LTRF2223-19-106]. X. Liu is supported by a Research Gift from Google. This research utilised Queen Mary's Apocrita HPC facility, supported by QMUL Research-IT. http://doi.org/10.5281/zenodo.438045. For the purpose of open access, the authors have applied a Creative Commons attribution (CC BY) license to any Author Accepted Manuscript version arising. The authors would like to thank the associate editor and the reviewers for their valuable comments to further improve this work.

\ifCLASSOPTIONcaptionsoff
  \newpage
\fi



\bibliographystyle{IEEEtran}
\bibliography{main}

\newpage
\appendices


\section{Audio-text alignment}\label{appendix:audio-text_alignment}
\begin{figure*}[t]
    \centering
    \includegraphics[scale=0.65]{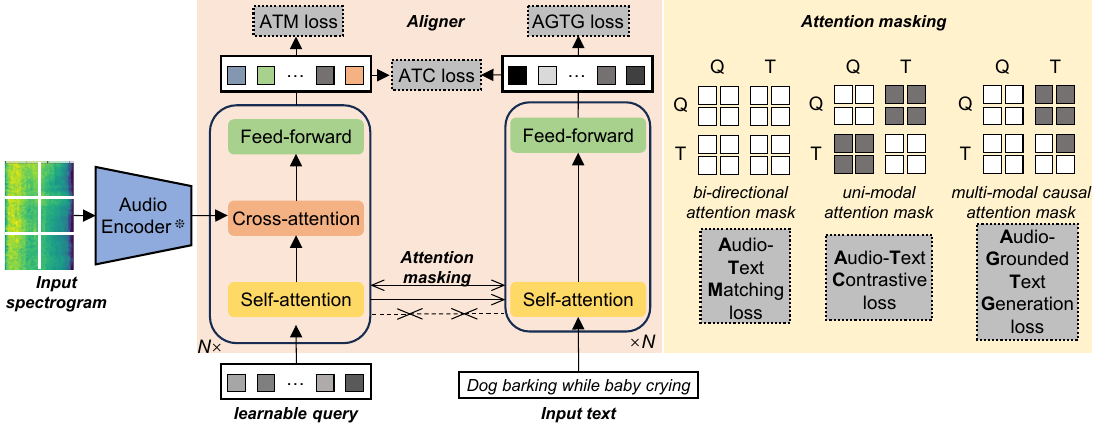}
    \caption{Illustration of APT audio aligner pretrained with audio-text pairs. The rest components are frozen in the pretraining. Using different self-attention masks, the parameter of the audio aligner is optimised with triplet learning objectives: Audio-Text Matching (ATM), Audio-Grounded Text Generation (AGTG), and Audio-Text Contrastive (ATC).}
    \label{fig.:pretraining}
\end{figure*}
Before coupling with an LLM, we pretrain the APT audio aligner to bridge the audio modality and the text modality. Fig.~\ref{fig.:pretraining} shows the implementation of audio-text alignment pretraining. Following Q-former pretraining in the previous work~\cite{li_blip-2_2023}, we freeze the other components and optimise the parameters of the audio aligner. During training, a fixed number of acoustic embeddings are learnt to extract relevant information from the audio feature maps according to the input text tokens. In the pretraining, the audio aligner learns with triplet training objectives:

\textbf{Audio-text matching (ATM).} The goal of ATM is to learn a token-level alignment between acoustic and text tokens. APT is trained to distinguish whether a pair of an audio recording and a text sentence are from the same source. A bi-directional self-attention mask is applied to the input sequence such that acoustic and textual tokens can attend to each other. APT's output embeddings $Z$ thus carry both audio and text information. Each vector of its embeddings $Z$ is then fed into a binary classifier, and the prediction is produced by averaging the binary logits over all vectors.

\textbf{Audio-grounded text generation (AGTG).} AGTG shapes APT's learning to generate texts, given the input audio tokens. The information for generating the text is first extracted from the audio tokens, and then passed to the text tokens via self-attention layers. In APT, a multimodal causal self-attention mask is used to control audio-text interaction, where each text token can attend to all audio tokens and its previous text tokens.

\textbf{Audio-text contrasting (ATC).} The ATC enforces alignment between the audio representation and the text representation such that their mutual information is maximized. It achieves so by contrasting the audio-text similarity of a positive pair against those of negative pairs. APT aligns acoustic embeddings output by the audio encoder with text tokens. To avoid information leakage, we employ a unimodal self-attention mask, where the acoustic and text tokens are not allowed to see each other.

\section{Prompts for various tasks}\label{appendix:multitask_prompt}
TABLE~\ref{tab.:proposed_task_template} detail the prompts we used to prompt APT-LLMs during training and inference. For each task, we hand-crafted three prompts as seeds and used ChatGPT to generate the rest of prompts. We checked the generated prompts and rejected those of ambiguity. We iterated the process till we got ten prompts per task.

\section{Analysis on few-shot audio classification}\label{appendix:few-shot}

\begin{table*}[h]
\centering
\caption{Template of audio tagging, automated audio captioning, language-queried sound event detection, temporal event retrieval,\\ and sound event counting.}
\label{tab.:proposed_task_template}
\resizebox{2\columnwidth}{!}{%
\begin{tabular}{@{}lllll@{}}
\toprule
Audio Tagging &
  Audio Captioning &
  \begin{tabular}[c]{@{}l@{}}Language-Queried Sound\\ Event Detection\end{tabular} &
  Temporal Event Retrieval &
  Sound Event Counting \\ \midrule
\begin{tabular}[c]{@{}l@{}}Summarize the audio with\\ key words.\end{tabular} &
  Summarize the audio succinctly. &
  \begin{tabular}[c]{@{}l@{}}Pinpoint the presence of\\ \{LABEL\} with the time\\ stamps.\end{tabular} &
  \begin{tabular}[c]{@{}l@{}}Summarize the audio with\\ key words in the interval\\ of \{STT\} seconds to \{EDT\}\\ seconds.\end{tabular} &
  \begin{tabular}[c]{@{}l@{}}How many times can the\\ sound \{LABEL\} be heard?\end{tabular} \\
\begin{tabular}[c]{@{}l@{}}What sound events can be\\ heard in the audio clip?\end{tabular} &
  \begin{tabular}[c]{@{}l@{}}Present a short overview of the\\ provided audio samples.\end{tabular} &
  \begin{tabular}[c]{@{}l@{}}Indicate the start and end\\ time of the audio event\\ \{LABEL\}.\end{tabular} &
  \begin{tabular}[c]{@{}l@{}}What sound events can be\\ heard from \{STT\} seconds\\ to \{EDT\} seconds?\end{tabular} &
  \begin{tabular}[c]{@{}l@{}}How many instances of\\ the sound \{LABEL\} can be\\ perceived?\end{tabular} \\
\begin{tabular}[c]{@{}l@{}}What auditory incidents can\\ be recognized in the recording?\end{tabular} &
  \begin{tabular}[c]{@{}l@{}}Provide a compact summary of\\ the auditory content.\end{tabular} &
  \begin{tabular}[c]{@{}l@{}}Document the exact times\\ the sound \{LABEL\} taking\\ place.\end{tabular} &
  \begin{tabular}[c]{@{}l@{}}What auditory incidents can\\ be recognized in the recording\\ from \{STT\} seconds to \{EDT\}\\ seconds?\end{tabular} &
  \begin{tabular}[c]{@{}l@{}}What is the number of\\ times the sound \{LABEL\} is\\ detectable?\end{tabular} \\
\begin{tabular}[c]{@{}l@{}}Which auditory occurrences\\ can be detected?\end{tabular} &
  \begin{tabular}[c]{@{}l@{}}Offer a brief outline of the audio\\ clips that have been given.\end{tabular} &
  \begin{tabular}[c]{@{}l@{}}Specify the time stamps for\\ \{LABEL\} occurrence.\end{tabular} &
  \begin{tabular}[c]{@{}l@{}}Which auditory occurrences\\ can be detected during \{STT\}\\ seconds to \{EDT\} seconds?\end{tabular} &
  \begin{tabular}[c]{@{}l@{}}How frequently can one\\ hear the sound \{LABEL\}?\end{tabular} \\
\begin{tabular}[c]{@{}l@{}}Which sound occurrences\\ can be perceived?\end{tabular} &
  \begin{tabular}[c]{@{}l@{}}Render a compressed version of\\ the audio’s main points.\end{tabular} &
  \begin{tabular}[c]{@{}l@{}}When the sound \{LABEL\}\\ happens?\end{tabular} &
  \begin{tabular}[c]{@{}l@{}}Which sound occurrences\\ can be perceived between\\ \{STT\} seconds and \{EDT\}\\ seconds?\end{tabular} &
  \begin{tabular}[c]{@{}l@{}}How often can the sound\\ \{LABEL\} be perceived?\end{tabular} \\
\begin{tabular}[c]{@{}l@{}}Present a concise breakdown\\ of the given audio clips.\end{tabular} &
  Describe the audio clip concisely. &
  \begin{tabular}[c]{@{}l@{}}Capture the exact times\\ when \{LABEL\} is\\ happening.\end{tabular} &
  \begin{tabular}[c]{@{}l@{}}Present a concise breakdown\\ of the recording from \{STT\}\\ seconds to \{EDT\} seconds.\end{tabular} &
   \\
\begin{tabular}[c]{@{}l@{}}List the sound events in the\\ audio clip.\end{tabular} &
  \begin{tabular}[c]{@{}l@{}}Explain the audio clip in a brief\\ and straightforward manner.\end{tabular} &
  \begin{tabular}[c]{@{}l@{}}Describe the time intervals\\ during which \{LABEL\}\\ takes place.\end{tabular} &
  \begin{tabular}[c]{@{}l@{}}List the sound events in the\\ interval of \{STT\} seconds to\\ \{EDT\} seconds.\end{tabular} &
   \\
\begin{tabular}[c]{@{}l@{}}Describe the recording with\\ names of sound events.\end{tabular} &
  \begin{tabular}[c]{@{}l@{}}Write a terse but informative\\ summary of the sound.\end{tabular} &
  \begin{tabular}[c]{@{}l@{}}State the precise moment\\ at which \{LABEL\} occurs.\end{tabular} &
  \begin{tabular}[c]{@{}l@{}}Name the auditory incidents\\ within the \{STT\} to \{EDT\}\\ seconds timeframe.\end{tabular} &
   \\
\begin{tabular}[c]{@{}l@{}}Enumerate the audio events\\ present in the audio.\end{tabular} &
  \begin{tabular}[c]{@{}l@{}}Give a quick overview of the\\ provided audio excerpts.\end{tabular} &
  \begin{tabular}[c]{@{}l@{}}What time does the sound\\ event \{LABEL\} take place?\end{tabular} &
  \begin{tabular}[c]{@{}l@{}}Enumerate the audio events\\ present between \{STT\} seconds\\ and \{EDT\} seconds.\end{tabular} &
   \\
\begin{tabular}[c]{@{}l@{}}Name the auditory incidents\\ in the audio sample.\end{tabular} &
  \begin{tabular}[c]{@{}l@{}}Outline the given audio samples\\ briefly.\end{tabular} &
  \begin{tabular}[c]{@{}l@{}}Capture the beginning and\\ end time of the sound\\ \{LABEL\}.\end{tabular} &
  \begin{tabular}[c]{@{}l@{}}Describe the recording with\\ names of sound events within\\ the \{STT\} to \{EDT\} seconds\\ timeframe.\end{tabular} &
   \\ \midrule
\#Output: \{LABEL\} &
  \#Output: \{CAPTION\} &
  \#Output: \{STT\}s-\{EDT\}s &
  \#Output: \{LABEL\} &
  \#Output: \{NUMBER\} \\ \bottomrule
\end{tabular}%
}
\vspace{-0.3cm}
\end{table*}

\begin{figure}[H]
    \centering
    \includegraphics[width=\columnwidth]{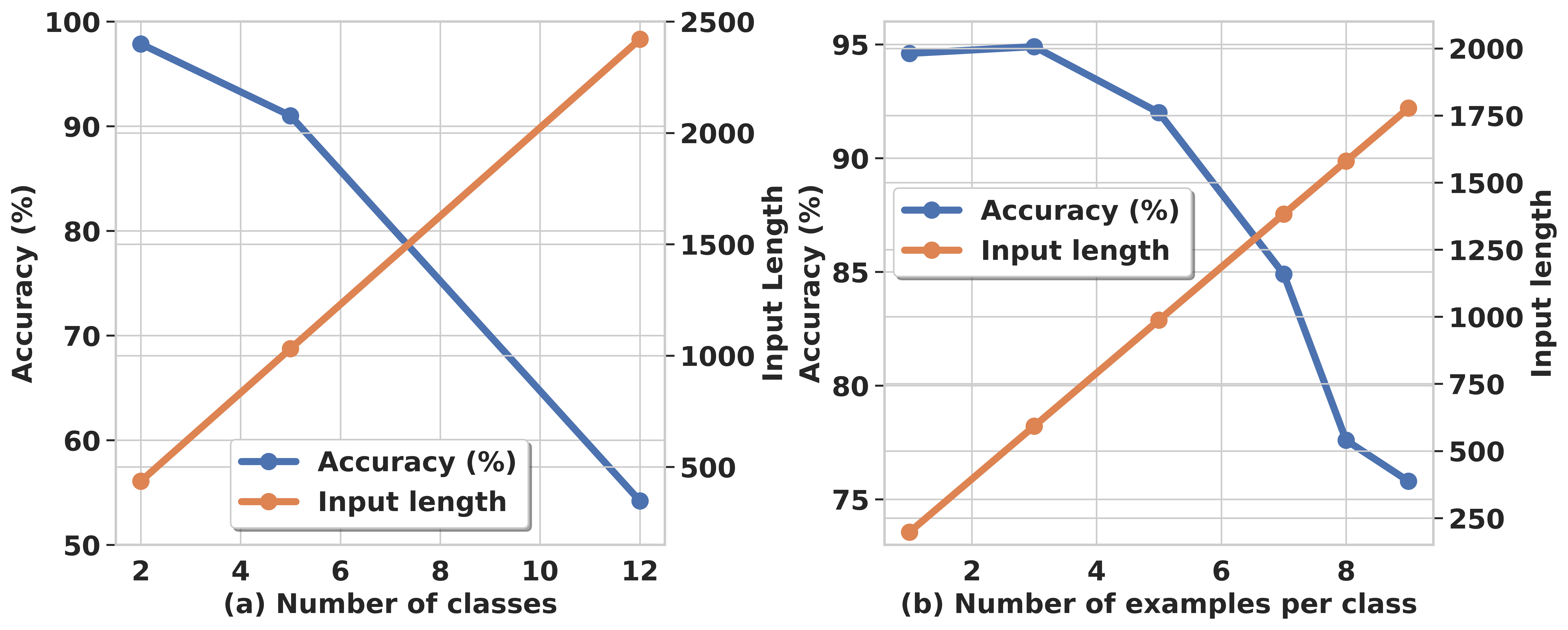}
    \vspace{-0.4cm}
    \caption{APT-LLM performance in few-shot audio classification when: (a) the number of classes increases, and (b) the number of examples per class increases. In both cases, the increase in the number of classes/examples leads to the rapid increase of the input sequence's length.}
    \label{fig.:few_shot}
\end{figure}

Figure~\ref{fig.:few_shot} (a) and (b) show APT-LLM performance in various few-shot settings. In both settings, the input sequence becomes longer when the number of shots/classes increases. It is interesting to observe that as the number of shots increases, the overall performance becomes worse. This is in contrast to the behavior of few-shot learners based on back propagation~\cite{liang_learning_2022}. One possible reason is that increasing the number of classes/shots rapidly increases the length of the input sequence, which is detrimental to APT-LLM global attention mechanism.

\end{document}